\documentclass[10pt]{IEEEtran}
\usepackage{pstricks}
\usepackage{pst-node}
\usepackage{pst-rel-points}
\usepackage{amssymb}
\usepackage{mathptm}
\usepackage{calc}
\usepackage{ifthen}
\usepackage{graphicx}

\newcommand{\bigsqcap}{ \begin{picture}(15,10)
                        \thicklines
                        \put(3,10){\line(1,0){9}}
                        \put(3,-1){\line(0,1){11}}
                        \put(12,-1){\line(0,1){11}}
                        \end{picture}
                      }
\newcommand{\MEETop}{\mbox{$\sqcap$}}
\newcommand{\NEWMEET}[2]{\mbox{\raisebox{-1.6ex}
                                {$\stackrel{#2}
                                {\stackrel{\bigsqcap}
                                        {\scriptstyle #1}}$}}}

\newenvironment{PROOF}{\noindent{\bf Proof :}}{}
\newtheorem{crlry}{Corollary}
\newenvironment{COROLLARY}{\crlry {\em\bf$\!\!\!\!$. }}{}
\newtheorem{thrm}{Theorem}
\newenvironment{THEOREM}[1]{\thrm \label{#1}{\em\bf$\!\!\!\!$. }}{}
\newtheorem{exmpl}{Example}
\newenvironment{EXAMPLE}[1]{\exmpl {\em }  \label{#1}\rm }{}

\newcommand{\edg}{{\em entity dependence graph}}
\newcommand{\EDG}{{\em EDG}}
\newcommand{\EDGs}{{\em EDG\/}s}
\newcommand{\nedg}{\mbox{$N_{\mbox{\scriptsize\em edg}}$}}
\newcommand{\eedg}{\mbox{$E_{\mbox{\scriptsize\em edg}}$}}
\newcommand{\en}{\mbox{$\alpha$}}

\newcommand{\Hfn}{\mbox{$\widehat{h}\,$}}
\newcommand{\weight}{\mbox{\sf\em Wt\/}}
\newcommand{\enNum}{\mbox{$\xi$}}
\newcommand{\depth}{\mbox{\sf\em d\/}}
\newcommand{\undef}{\mbox{\sf\em undef\/}}
\newcommand{\nonconst}{\mbox{\sf\em nonconst\/}}
\newcommand{\fmod}{\mbox{\sf\em dfpmod$\!\,_f$\/}}
\newcommand{\fuse}{\mbox{\sf\em dfpuse$_f$\/}}

\newcommand{\LD}{\mbox{${L}$}}
\newcommand{\topd}{\mbox{$\top$}}
\newcommand{\botd}{\mbox{$\bot$}}
\newcommand{\Lt}{\mbox{$\widehat{L}\,$}}
\newcommand{\Tt}{\mbox{$\widehat{\top}$}}
\newcommand{\Bt}{\mbox{$\widehat{\bot}$}}
\newcommand{\Xt}{\mbox{$\widehat{X}$}}
\newcommand{\Yt}{\mbox{$\widehat{Y}$}}

\newcommand{\HL}{\mbox{$H$}}

\newcommand{\Ht}{\mbox{$\widehat{H}$}}
\newcommand{\dff}{\mbox{\boldmath$D$}}
\newcommand{\idff}{\mbox{\boldmath$I_D$}}
\newcommand{\CFG}{\mbox{CFG}}

\begin{document}

\title{Complexity of Data Flow Analysis for Non-Separable Frameworks}

\author{\begin{tabular}{cc}
	Bageshri (Sathe) Karkare$^\ast$\thanks{$^\ast$Supported by Infosys 
	  Technologies Limited, Bangalore, under \mbox{Infosys\/} Fellowship
	  Award.}\ \  & Uday Khedker \\
	{Dept. of Computer Science \& Engg.}, &{Dept. of Computer Science \& Engg.}, \\ 
	Indian Institute of Technology,  Bombay &
	Indian Institute of Technology,  Bombay \\
        Mumbai, India & Mumbai, India  \\
	{\tt bageshri@cse.iitb.ac.in} & 
	{\tt uday@cse.iitb.ac.in} 
	\end{tabular}}
\date{}
\maketitle
\thispagestyle{empty}

\begin{abstract}
\normalsize \rm 
{\em The  complexity of round  robin iterative data flow  analysis has
been traditionally defined as \mbox{$1 + \depth$} where \depth\ is the
depth of a  control flow graph.  However, this  bound is restricted to
bit  vector  frameworks,  which  by definition,  are  separable.   For
non-separable frameworks, the complexity  of analysis is influenced by
the interdependences of program entities, hence the bound of
\mbox{$1+\depth$}  is not  applicable.   This motivates  the need  for
capturing the interdependences of  entities to define a general
complexity measure.

We propose  {\em Degree of  dependence} $\delta$ which  quantifies the
effect  of  non-separability  on  the  complexity of  analysis  for  a
particular problem instance. We define the complexity bound of
\mbox{$1+\delta+\depth$} which explains the complexity of round
robin analysis of general non-separable data flow problems.  Like
\depth,  $\delta$  is  a theoretical  concept useful  for
understanding the complexity rather  than estimating it. In bit
vector  frameworks  the   bound  \mbox{$1+\delta+\depth$}  reduces  to
\mbox{$1+\depth$}  due  to  $\delta  =  0$.   Apart  from  being
general,  our bound  is  also precise,  as  corroborated by  empirical
results. }

\end{abstract}


\section{Introduction}
\label{sec:intro}

For  a given  instance of  a data  flow framework,  the  complexity of
performing  data  flow  analysis  depends  on  a  combination  of  the
properties  of  program  structure  (in particular,  loops  and  their
complex arrangements), separability and boundedness of flow functions,
height of  the lattice,  and the order  of traversal over  the control
flow  graph.  Among  the  general complexity  measures  for data  flow
analysis (\cite{asu, width, hecht-book, dfa-chapter, gen-theory-bvdfa,
marlowe-properties,   muth-complexity-dfa,   nielson-bounded-fp})  the
number of iterations required for convergence of round robin iterative
method has been explored in details.  The traditional complexity bound
of round robin  data flow analysis for bit  vector frameworks has been
defined   as   $1  +   \depth$   iterations  (\cite{asu,   hecht-book,
dfa-chapter}), where  \depth\ is the  depth of the control  flow graph
which is  defined as the maximum  number of back edges  in any acyclic
path\footnote{For bidirectional frameworks,  this has been generalized
to $1 + w$  (\cite{width, dfa-chapter, gen-theory-bvdfa}) where $w$ is
the maximum width of any information flow path.}.

Depth $\depth$  is not  independent of the  problem instance  since it
captures the structure of control flow. However, it does not depend on
the  program  statements.  Since  bit-vector  problems  are  separable
$2$-bounded  problems, program  statements do  not play  a significant
role  in increasing  the  complexity. Hence  their  complexity can  be
reasonably explained  using only the depth  $\depth$. In non-separable
frameworks   like   constant   propagation   (\cite{asu,   hecht-book,
muchnick-book}), faint  variables analysis (\cite{hor-fv, knoop-pde}),
type  inferencing of flow  sensitive types~\cite{kdm.typeinferencing},
and points-to  analysis (\cite{emami-ip-points-to, aditya-hetero}) the
complexity  of analysis  is influenced  by the  interdependences among
program  entities,  which  in  turn  are  heavily  influenced  by  the
placement  of  program statements,  apart  from  depth. Complexity  of
analyzing such problems has been observed to be proportional to number
of program entities, but no strict bound has been reported.

We   define  an  {\em   Entity  Dependence   Graph}  to   capture  the
non-separable  information flows  in the  given program  instance.  We
propose  $\delta$, the  {\em degree  of  dependence} as  a measure  of
effect of non-separability on complexity.  We use $\delta$ to define a
complexity  bound  of  $1+\delta+\depth$   which  is  the  first  ever
realistic  explanation of  the complexity  of data  flow  analysis for
non-separable  frameworks.  Note that  like  $\depth$,  $\delta$ is  a
purely theoretical  concept which is used for  explanation rather than
for  estimation  of  complexity.  Our complexity  bound  is  uniformly
applicable to  a large class  of data flow frameworks.  In particular,
the separability of  bit vector frameworks can be  viewed as a special
case of non-separability, due to which $\delta = 0$ and the bound $1 +
\delta + \depth$  reduces to $1 + \depth$.   Apart from being general,
our  measure  is  precise.  Our empirical  measurements  for  constant
propagation and faint variables analysis corroborate this.

The rest of the paper is organized as follows: Section~\ref{sec:dff}
reviews the theory of data flow analysis.  
Section~\ref{sec:edg} introduces the entity dependence
graph. Section~\ref{sub:edg.paths}  defines the degree of
dependence and derives the bound of
$1+\delta+\depth$. Section~\ref{sec:data}  provides empirical
measurements.

\section{Data Flow Frameworks}
\label{sec:dff}

A      data     flow      framework     \dff\      is      a     tuple
\mbox{$\langle\LD,\sqcap,F\rangle$}    (\cite{asu,    graham-fastalgo,
hecht-book,  ku,  dfa-chapter, kildall-unified,muchnick-book}),  where
\LD\  is a  lattice with  meet $\sqcap$  and $F$  is the  set  of flow
functions  \mbox{$\LD \mapsto\LD$}.   \LD\  is the  set  of data  flow
values with  the partial order  $\sqsubseteq$ induced by  $\sqcap$ and
has  a top  element \topd\  and a  bottom element  \botd.  An instance
\idff,  of  a data  flow  framework  \dff,  is a  pair  \mbox{$\langle
G,M\rangle$} where $G$ is a {control} flow graph {(\CFG)} and $M$ is a
mapping  from  the nodes/edges  of  the  {\CFG}  to the  functions  in
$F$. Note that \dff\ defines only the structure of $L$, it is actually
instantiated by  \idff\ depending upon  the actual number  of entities
(eg. variables or expressions etc.).

\subsection{Properties of Lattices}

For a lattice  $L$, a {\em strictly descending  chain\/} is a sequence
\mbox{$v_0 \sqsupset v_1 \sqsupset  \ldots \sqsupset v_{n}$} such that
$v_i  \in L$.   Analogously,  a {\em  strictly  ascending chain\/}  is
\mbox{$v_n  \sqsubset v_{n-1}  \sqsubset  \ldots \sqsubset  v_{0}$}.
The lattices in  which all strictly descending chains  are finite have
been  called  {\em bounded}  in  (\cite{ku,kildall-unified}) and  {\em
finite\/} in~\cite{hecht-book}.   The lattices in  which both strictly
descending as well  as strictly ascending chains are  finite have been
called  {\em  finite\/}  in (\cite{dfa-chapter,gen-theory-bvdfa})  and
{\em complete\/} in (\cite{nielson-flow-logics}).  In the rest of this
paper, by bounded lattices we mean complete lattices.

The  overall  lattice  $\LD$   is  a  product  of  component  lattices
\mbox{$\Lt_1          \times          \Lt_2\times         \cdots\times
\Lt_{\scriptsize\enNum}$}, where $\Lt_i$ is a lattice of the values of
data flow  properties of an individual entity  {$\en_i$}, and $\enNum$
is the  number of entities.   The \mbox{$\top, \bot\in L$}  are tuples
\mbox{$\langle\Tt_1 ,  \Tt_2, \ldots, \Tt_{\scriptsize\enNum}\rangle$}
and         \mbox{$\langle\Bt_1        ,         \Bt_2,        \ldots,
\Bt_{\scriptsize\enNum}\rangle$}.
In the case of available  expressions analysis, for a given expression
$e$,  $\Tt$  is  $\{e\}$   and  $\Bt$  is  $\emptyset$.  $\sqcap$  and
$\sqsubseteq$ are $\cap$ and  $\subseteq$ respectively. In the case of
reaching definitions  analysis, for a given  definition $di :  x = e$,
$\Tt$  is $\emptyset$  and $\Bt$  is $\{di\}$.  Further,  $\sqcap$ and
$\sqsubseteq$ are $\cup$ and $\supseteq$ respectively.

The  height of  a  lattice is  the  maximum number  of $\sqsubset$  or
$\sqsupset$ in any strict chain. Let $\Ht_i$ denote height of $\Lt_i$.
Often all $\Lt_i$ are same and the height $\HL$ of the overall lattice
$\LD$ is:
\begin{equation}
\HL = \Ht_i \times \enNum
\label{eq:lattice.height}
\end{equation}

\subsection{Properties of Flow Functions}

Flow functions are {\em
monotonic\/} (\cite{asu, graham-fastalgo, hecht-book, ku, dfa-chapter,
  kildall-unified,muchnick-book}) if:
\begin{equation}
\forall f \in F, \forall x,y \in L\;: x \sqsubseteq y \Rightarrow f(x)
\sqsubseteq f(y)
\label{eq:monotonicity}
\end{equation}

If the  flow functions \mbox{$f : L  \mapsto L$} in $F$  are tuples of
functions \mbox{$f = \langle \Hfn_1,\Hfn_2,\cdots,\Hfn_k\rangle$} such
that \mbox{$\Hfn_i  : \Lt_i\mapsto\Lt_i$}, then the  framework is {\em
separable\/}   (\cite{dfa-chapter,gen-theory-bvdfa})   in   that   the
functions on one component lattice operates independently of others.

The  flow  functions  are  {\em  $k$-bounded\/}  if  their  {\em  loop
closures\/}    (\cite{dfa-chapter,gen-theory-bvdfa,marlowe-properties})
(also       called       {\em       fastness      closures\/}       in
\cite{graham-fastalgo,hecht-book,nielson-flow-logics})  are bounded by
constant $k$. Let $f^{j+1} = f\circ f^{j}$ and $f^{0}$ be the identity
function. $f^j$ represents the flow function corresponding to the path
containing $j$ traversals over a loop.  $k$-boundedness requires that
\begin{equation}
\exists \;k \geq 1 \; s.t. \; \forall \; f \in F: \; f^{0} \MEETop
f^{1} \MEETop f^{2} \MEETop \cdots = \NEWMEET{i=0}{k-1} f^{i}
\label{eq:loop.closures}
\end{equation}
$k$-boundedness implies that though  a program contains infinite paths
in the  presence of  loops, a finite  number of paths  containing upto
$k-1$ unfoldings of loops are  sufficient for convergence of data flow
analysis.

The computation  of \mbox{$f^{0}  \MEETop f^{1} \MEETop  f^{2} \MEETop
\cdots \MEETop f^{i}$} represents the {\em greatest lower bound} ({\em
glb}) of the results of first $i$ applications of $f$. Thus it follows
a descending chain  in \LD. Hence bounded lattices  imply bounded loop
closures.

\begin{figure}[t]
\scalebox{.9}{\includegraphics{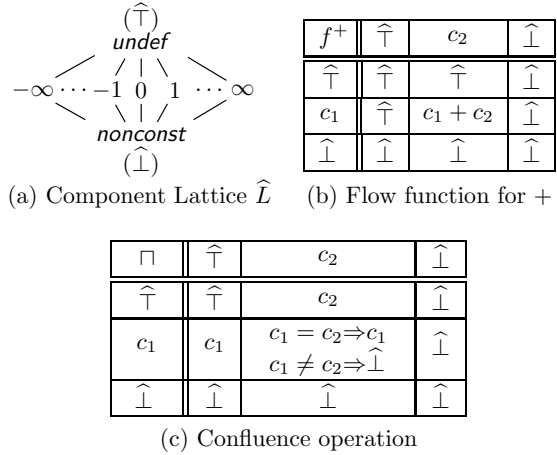}}
\caption{Constant propagation framework.}
\label{fig:cp.lattice}
\rule{\columnwidth}{.7pt}
\end{figure}

For separable frameworks, values  in \Lt\ change simultaneously due to
independence of the component lattices.  It is easy to verify that for
a $k$-bounded framework,
\begin{equation}
k   \leq  \left\{ \begin{array}{ll}
		\Ht + 1 & \mbox{ if the framework is separable} \\           
		\HL + 1 & \mbox{ if the framework is non-separable}            
		\end{array}
		\right.	\label{eq:loop.closure.height}
\end{equation}

Bit vector  frameworks have separable flow functions  which operate on
\Lt\  with \mbox{$\Ht=1$}. Further,  since all  flow functions  in bit
vector   frameworks  are  either   constant  functions   or  identity,
\mbox{$f^2  (x) =  f(x)$} implying  that $k  $ is  $2$ for  bit vector
frameworks.

\subsection{Examples of Non-Separable Frameworks}
\label{app:cp}

In this  section, we briefly introduce constant  propagation and faint
variables analysis which are used as running examples in the paper.

{\em     Constant     propagation}~\cite{asu,hecht-book,muchnick-book}
identifies variables  which hold a  fixed constant value  and replaces
them by this value.  Lattice $\Lt$  of data flow values for a variable
is  shown  in   Figure~\ref{fig:cp.lattice}(a).   \undef\  denotes  an
undefined  value  and forms  the  \Tt\  whereas  \nonconst\ denotes  a
non-constant  value and  forms  the \Bt.   All constants  \mbox{$c_i$,
$-\infty  \leq c_i  \leq  \infty$} are  ordered  such that  \mbox{$c_i
\sqsubset \Tt$ and $\Bt \sqsubset c_i$}. Observe that this lattice has
an infinite number  of elements but a finite  height: \mbox{$\Ht = 2$}
and \mbox{$\HL  = 2\times\enNum$}.  Confluence  operation for constant
propagation   is   shown   in  Figure~\ref{fig:cp.lattice}(c).    Flow
functions  which  influence  the  data  flow  properties  of  variable
correspond to  statements such  as $x=y$, $x=  y+z$, $x  = 3$ or  $x =
read()$.  For the first two statements, the flow functions compute the
data flow  value of  $x$ from  the data flow  values of  the variables
appearing in  the right hand  side.  For the  last two, the  data flow
values    of   $x$   are    $c_3$   and    $\nonconst$   respectively.
Figure~\ref{fig:cp.lattice}(b)  shows the flow  function corresponding
to a statement $x = y+z$. Observe that it is of the form \mbox{$(\Lt_x
\times \Lt_y) \mapsto  \Lt_z$} rather than \mbox{$\Lt\mapsto\Lt$}. Due
to non-separability,  its loop closure bound  $k$ is $H+1$  and is not
constant.

{\em  Faint  variables  analysis}~\cite{hor-fv,knoop-pde}  is  a  more
general variant  of live  variables analysis in  that it  computes the
transitive closure  of dead variables.  A  variable is faint  if it is
dead or if  it is only used to compute new  values of faint variables.
Variables which are not faint are {\em strongly live}~\cite{NNH99fe}.

\begin{figure}[t]
\scalebox{.9}{\includegraphics{fig_fv.epsi}}
\caption{Data flow equations to determine faint variables}
\label{fig:dfeqn.faint}
\rule{\columnwidth}{.7pt}
\end{figure}

Data  flow  equations  to   identify  faint  variables  are  given  in
Fig.~\ref{fig:dfeqn.faint}.    For   a    variable   $v$,   $\Tt$   is
\mbox{$\{v\}$}, $\Bt$  is $\emptyset$ and $\Ht=1$.   Though these data
flow  properties  can  be  represented  using bit  vectors,  the  flow
functions are non-separable. Hence this is not a bit vector framework.

\renewcommand{\a}[2]{\mbox{$\alpha_{\scriptsize#1}^{\scriptsize#2}$}}
\renewcommand{\r}{\mbox{$\;-\!\!\triangleright\;$}}
\newcommand{\rn}[2]{\rnode{#1}{#2}}
\newcommand{\sHt}{\mbox{$\scriptsize\Ht$}}

\subsection{Simplistic Generalization of Complexity Bound}

Using only the properties of framework, a simplistic generalization of
the  complexity bound  can be  formulated as  follows: In  round robin
method, in the worst case,  each iteration may change exactly one data
flow  value. For general  frameworks, at  most $H$  new values  may be
generated at  any program point.  Since computation of each  new value
may require at most \depth\ additional iterations, the total number of
iterations  is  \mbox{$1 +  (\depth  \times  H)$}.   In a  $k$-bounded
framework, at  most $(k-1)$ new values  can be generated  at a program
point. Hence $(k-1)$ can replace  $H$ giving a possibly stricter bound
of  $1+\depth\times(k-1)$ (see Equation~\ref{eq:loop.closure.height}).
This  has already  been  conjectured in~\cite{dfa-chapter}.   However,
since $k$ may  not be known, we continue  to use $1+(\depth\times H)$.
For separable frameworks, it is $1+(\depth \times \Ht)$.

\begin{figure}[t]
\scalebox{.9}{
\begin{pspicture}(0,0)(7.5,9.4)
\psset{unit=1mm} 
\psrelpoint{origin}{n1}{65}{80}
\rput(\x{n1},\y{n1}){\rnode{n1}{$1\;$\psframebox[framearc=.2]{$w =
      2$}$\;\white 1$}} 
\psrelpoint{n1}{n2}{0}{-10} 
\rput(\x{n2},\y{n2}){\rnode{n2}{$2\;$\psframebox[framearc=.2]{$
      \mbox{\em print } x$}$\;\white 2$}} 
\psrelpoint{n2}{n3}{0}{-10} 
\rput(\x{n3},\y{n3}){\rnode{n3}{$3\;$\psframebox[framearc=.2]{$\white
      w = 2$}$\;\white 3$}} 
\psrelpoint{n3}{n4}{0}{-9} 
\rput(\x{n4},\y{n4}){\rnode{n4}{$4\;$\psframebox[framearc=.2]{$\white
      x = 2$}$\;\white 4$}} 
\psrelpoint{n4}{n5}{0}{-9} 
\rput(\x{n5},\y{n5}){\rnode{n5}{$5\;$\psframebox[framearc=.2]{$x =
      y\!+\!2$}$\;\white 5$}} 
\psrelpoint{n5}{n6}{0}{-9} 
\rput(\x{n6},\y{n6}){\rnode{n6}{$6\;$\psframebox[framearc=.2]{$y =
      z\!+\!3$}$\;\white 6$}} 
\psrelpoint{n6}{n7}{0}{-9} 
\rput(\x{n7},\y{n7}){\rnode{n7}{$7\;$\psframebox[framearc=.2]{$z =
      w\!-\!1$}$\;\white 7$}} 
\psrelpoint{n7}{n8}{0}{-9} 
\rput(\x{n8},\y{n8}){\rnode{n8}{$8\;$\psframebox[framearc=.2]{$w =
      x\!+\!1$}$\;\white 8$}} 
\ncline{->}{n1}{n2} 
\ncline{->}{n2}{n3} 
\ncline{->}{n3}{n4} 
\ncline{->}{n4}{n5} 
\ncline{->}{n5}{n6} 
\ncline{->}{n6}{n7} 
\ncline{->}{n7}{n8} 
\ncarc[arcangle=235,ncurv=.8]{->}{n8}{n4}
\ncarc[arcangle=130,ncurv=2.2]{->}{n4}{n3}
\ncarc[arcangle=230,ncurv=1.7]{->}{n3}{n2}
\rput[lt](-5,95){\begin{tabular}[t]{l@{\ \ \ \ }l}
		\multicolumn{2}{l}{\em Constant Propagation} \\
		- & $w$ is 2 at the exit of 4.  \\
		  & No variable is constant\\
                  & at any other point.\\
		- & $\Ht = 2,\; \HL = 8, \;w = 3 $\\
		- & Predicted bound : 25 iterations \\
		- & Actual iterations : 9\\ 
                - & If the assignments in nodes 5\\
		  & and 7 are exchanged, \Ht, \HL,\\
		  & and $\depth$ (hence predicted bound) \\ 
		  & remain same. Actual number of \\
		  & iterations reduces to 5. \\ \\
		\multicolumn{2}{l}{\em Faint Variables Analysis} \\
		  - & Due to the {\em print\/} statement, no\\
		    & variable is faint at any point.\\
		- & $\Ht = 1,\; \HL = 4,\;w = 3 $\\
		- & Predicted bound : 13 iterations \\
		- & Actual iterations : 7\\
                - & If the assignments in nodes 5 and 7 \\
		  & are exchanged, \Ht, \HL, and $\depth$ \\
		  & (hence predicted bound) remain same.\\
                  & Actual number of iterations reduces to 5.
		\end{tabular}
		}
\end{pspicture} 
}
\caption{Complexity bounds for non-separable analyses are very loose.}
\label{fig:existing.bound.lim}
\rule{3in}{0.7pt}
\end{figure}

\begin{EXAMPLE}{}\label{exmp:cp.fva.iterations}%
Consider  constant propagation  and faint  variables analysis  for the
\CFG\ in Figure~\ref{fig:existing.bound.lim}.  Observe that the actual
number  of  iterations  is  much  smaller than  the  predicted  bound.
Besides, a rearrangement of  statements causes a significant variation
in the number of iterations  and the predicted bound is insensitive to
this change.
\end{EXAMPLE}

\ 

Separability guarantees that the  data flow properties of all entities
are  independent  of  each  other  whereas  handling  non-separability
requires the assumption that in the worst case, the data flow property
of  every  entity   depends  on  the  data  flow   properties  of  all
entities.  We  view these  two  extreme  cases  of dependence  as  two
extremes of the  same continuous spectrum by modeling  the exact cause
of  non-separability   explicitly  and  by  defining   the  degree  of
dependence as a measure  of non-separability. The degree of dependence
is defined  for a  particular problem instance  using the  {\em Entity
Dependence Graph} which we introduce in the next section.

\section{Entity Dependence Graph}
\label{sec:edg}

Let the  flow function corresponding to  a statement $s$  be $f$.  Let
\fmod\ denote  the set of entities  which occur in $s$  and whose data
flow  properties  are computed  by  $f$. Note  that  for  the flow  of
information, this  computation should potentially  compute a non-$\Tt$
value  because  information  flow  consists of  propagating  non-$\Tt$
value.  Let \fuse\  denote the set of entities which  occur in $s$ and
whose data flow properties are used by $f$.

Formally, an \edg\ (\EDG) is a directed graph $G_{\mbox{\scriptsize\em
edg}} = \langle \nedg, \eedg \rangle$.
\nedg\ is {the set of entities} defined as follows:

$\nedg = \{ \en_s \mid \en \in \fmod, \; f$ is a flow function for
statement $s\}$

\en\ is the name of an entity whereas $\en_s$ represents its
instantiation for statement $s$. $\en_s^v$ denotes entity $\en_s$ with
value  $v$.  Different instances  of the  same entity  may have  to be
created in the following situations:
\begin{enumerate}
\item When a data flow problem requires such a distinction. For
      example, reaching definitions analysis requires enumeration of
      all definitions of each variable reaching a program point.
\item When different instances of the same entity reaching a program
      point have different dependence relations and hence different
      influences on the complexity.
\end{enumerate}

\noindent \eedg\ is the set of edges between entities.
An  edge \mbox{$\en_i  \rightarrow \beta_j$}  indicates that  the data
flow property  of entity $\beta_j$  directly depends on the  data flow
property of  $\en_i$, where $i$ and  $j$ may not be  adjacent in \CFG.
$\en_i^v \rightarrow \beta_j^u$ indicates  that value $u$ of $\beta_j$
is due to the direct  influence of value $v$ of $\en_i$.  Constructing
a dependence \mbox{$\en_i \rightarrow \beta_j$} requires:

\begin{itemize}
\item identifying  the presence of a flow function $f$ at a
      program point $p$ associated with statement $j$ such that
      \mbox{$\beta_j\in\fmod$} and \mbox{$\en\in\fuse$}, and
\item discovering the instance $\en_i$ reaching $s$ for
      \mbox{$\en\in\fuse$}.
\end{itemize}      

For information  propagation, each \EDG\ must contain  some entry node
i.e. a node with no predecessors: Some entity must change from
\Tt\ to non-\Tt\ independently, otherwise all
entities would remain \Tt. In the latter situation, data flow analysis
need not be performed.  For separable frameworks EDG-edges can exist
only between instances of the same entity because data flow properties
of an entity change independently of other entities.

\begin{figure}[t]
\includegraphics{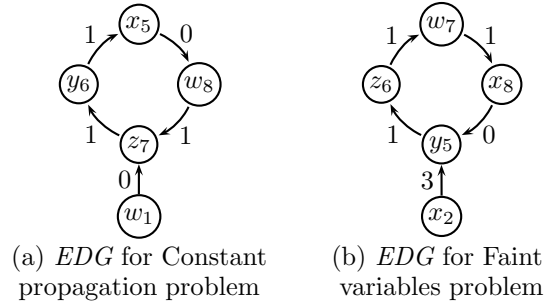}
\caption{Entity Dependence Graphs for the \CFG\ in
  Figure~\protect\ref{fig:existing.bound.lim}. Edges are labeled
  with edge-weights.}
\label{fig:example.edgs}
\raisebox{-.2cm}{\rule{\columnwidth}{.7pt}}
\end{figure}

\begin{EXAMPLE}{}
Some \EDGs\ for  the \CFG\ in Figure~\ref{fig:existing.bound.lim} 
are as described below:

{\em Available Expressions Analysis:} In a bit vector framework, flow
      functions are {either} constant functions {or identity} and all
      entities are independent of others. Hence \EDG\ has no
      edges. For available expressions analysis, an entity $\alpha_s$
      represents the fact that expression $\alpha$ is killed in
      statement $s$. For the instance in
      Figure~\ref{fig:existing.bound.lim}, the set of \EDG-nodes is
      {\mbox{$\{(y+2)_6,(z+3)_7,(w-1)_1,(w-1)_8,(x+1)_5\}$}} and the
      set of edges is $\emptyset$.

{\em Constant Propagation:} For constant propagation, an entity is
      identified by a variable name and statement number associated
      with a definition of that variable. The resulting \EDG\ is shown
      in Fig.~\ref{fig:example.edgs}(a). Nodes $w_8$ and $w_1$
      correspond to definitions of $w$ in statements $8$ and $1$
      respectively. A reaching definitions analysis using the renamed
      definitions can identify the \EDG-edges.  An edge label
      represents weight of an edge (defined in
      Section~\ref{sub:edg.paths}).

{\em Faint Variables Analysis:} Figure~\ref{fig:example.edgs}(b) shows
      the \EDG.  The uses of $x$ in statements $2$ and $8$ are treated
      as different entities denoted by $x_2$ and $x_8$ respectively.
      An assignment \mbox{$s :a=b$} adds dependence edges
      \mbox{$a_i\rightarrow b_s$} such that $a_i$ is live at exit of
      $s$. Identifying dependence edges requires live variables
      analysis with renamed uses.
\end{EXAMPLE}

\ 

An \EDG-path $\alpha_i\r\beta_j$  represents a transitive influence of
$\alpha_i$    on    $\beta_j$.      Cycles    in    \EDG\    represent
self-dependences. Interestingly, a cyclic  \EDG-path may differ from a
cyclic \CFG-path.  For example, in the \EDG\  for constant propagation
in Figure~\ref{fig:example.edgs}(a), path  \mbox{$(z_7, y_6, x_5, w_8,
w_7)$}  captures the  cyclic dependence  among entities  $z_7$, $y_6$,
$x_5$,      $w_8$,       and      $z_7$.       The       \CFG\      in
Figure~\ref{fig:existing.bound.lim}    contains   many    cycles   not
necessarily coinciding with this \EDG-cycle. Further, each edge in the
\EDG-cycle corresponds to a cyclic path involving the back edge
\mbox{$8\rightarrow 5$} in the \CFG.

Observe that  the dependence captured  by \EDG\ is different  from the
dependences  explored in  past.  Control dependence\cite{ferrante-pdg}
captures  the  dependence  of   execution  of  a  statement  on  other
statements.   Data  dependences\cite{kuck-dg},  Def-Use   and  Use-Def
chains\cite{asu},   SSA   edges\cite{cytron-ssa}   etc.  capture   the
dependence of  a statement on  the data computed in  other statements.
These traditional  dependences do not  capture the dependence  of data
flow properties and hence cannot be used for complexity analysis.

\section{Defining Complexity Using The Degree of Dependence}
\label{sub:edg.paths}

For   an  \EDG\  edge   \mbox{$\en_i\rightarrow\beta_j$},  edge-weight
denoted  as $\weight(\en_i  \rightarrow  \beta_j)$ is  defined as  the
maximum number of back edges in any acyclic path from stmt $i$ to stmt
$j$ for forward  problems, and from stmt $j$ to  stmt $i$ for backward
problems\footnote{for  bidirectional  problems,  the  notion  of  {\em
width}~\cite{gen-theory-bvdfa} must be used.}.

\begin{EXAMPLE}{}
All edges in the \EDG\ shown in Figure~\ref{fig:example.edgs} are
labeled with their weights.  In the \EDG\ for faint variables
analysis, edge $x_2 \rightarrow y_5$ has weight $3$ since the path
from stmt $5$ to stmt $2$ contains three back-edges.  Edge \mbox{$x_8
\rightarrow y_5$} has depth $0$ because of the back-edge-free path
from stmt $5$ to stmt $8$.
\end{EXAMPLE}

\ 

\newcommand{\edgpath}{\mbox{$\alpha_i -\!\!\!\triangleright \beta_j$}}
\newcommand{\dd}[1]{\mbox{$\Delta(#1)$}}

The {\em weight\/} of \EDG-path \edgpath\ denoted
\mbox{$\weight(\edgpath)$}, is defined as the sum of weights of the
edges along the path.  It is defined only for an acyclic path with the
relaxation that $\alpha_i$ and $\beta_j$ may be same.  The {\em degree
of dependence} of an \EDG-path \edgpath, denoted
\dd{\edgpath}\ is defined according to the structure of the path as
follows: 

\begin{itemize}
\item If the path \edgpath\  is  acyclic, $\dd{\edgpath} =
      \weight(\edgpath)$. 
\item If the path \edgpath\ is acyclic except that $\beta_j$ is same as
      $\en_i$, then a change in $\en_i$ has to be propagated to all
      entities along the path, including $\en_i$. Doing so once may
      require $\weight(\edgpath)$ additional iterations. This may
      change $\en_i$ further. The maximum number of changes is \Ht,
      and 
      \begin{eqnarray}
      \dd{\edgpath}& = &  \Ht\times\weight(\edgpath)
	\label{eq:path.cyclic}
      \end{eqnarray}
      {In case of overlapping cycles, the maximum  value of
      $\dd{\en_i\r\en_i}$ is considered.} 

\item If the path \edgpath contains $m$ non-overlapping cycles
      connected by $m-1$ acyclic segments, it has the
      following structure:
	\begin{eqnarray}
	\en_i\r
	\gamma_1\r\gamma_1\r
	\gamma_2\r\gamma_2\r
	\cdots
	\gamma_m\r\gamma_m\r
	\beta_j
	\label{eq:path.structure}
	\end{eqnarray}
	Then $\dd{\edgpath}$ is defined as:
      \begin{eqnarray}
      \!\!\!\!\!\!\!\!\!\!\!\dd{\edgpath}& = & 
      {\displaystyle\sum_{n=1}^{m}}
      {\dd{\gamma_n\r\gamma_n}} 
      +{\displaystyle\sum_{n=1}^{m-1}}
      {\dd{\gamma_n\r\gamma_{n+1}}}\nonumber\\
      & &   
      + {\dd{\en_i\r\gamma_1}}
       + \dd{\gamma_m\r\beta_j} 
      \label{eq:Delta.non-overlapping}
      \end{eqnarray}
\end{itemize}

$\dd{\en_i\r\beta_j}$  computes  the   maximum  number  of  iterations
required to  propagate the influence of  a change in the  value of the
data flow property  of $\en_i$ on the value of  the data flow property
of $\beta_j$ along the path.  Let  $\en_0$ denote an entry node of the
\EDG\  for  an  instance \idff\  for  a  given  \dff.  The  degree  of
dependence of \idff\ is denoted by $\delta$ and is defined as follows:
\begin{eqnarray}
\delta = \max(\dd{\en_0\r\beta_j}), \mbox{ for any $\en_0$ and
$\beta_j$} 
\label{eq:degree.def}
\end{eqnarray}

In  the  case  of   practical  non-separable  problems  like  constant
propagation and  faint variables analysis,  $\delta$ can be  made more
precise. We define the  entity dependence in a non-separable framework
to be monotonic if
\begin{equation}
\forall \en_i^v\rightarrow \beta_j^u ,\; ht(u) \geq ht(v)
\label{cond:monotonicity.entity}
\end{equation}

$ht(v)$  denotes the height  of $v$  in the  component lattice  and is
defined as the length of a longest descending chain from $\Tt$ to $v$.
For   \mbox{$Y=f(X)$},  let   \mbox{$X,Y\in   L$}  be   \mbox{$\langle
\Xt_1,\Xt_2,\ldots,\Xt_{\scriptsize\enNum}\rangle$} and
\mbox{$\langle \Yt_1,\Yt_2,\ldots,\Yt_{\scriptsize\enNum}\rangle$}.
Let $\Xt_i$ determine the value of $\Yt\!_j$.
Then the possible values of $\Yt\!_j$ are limited to those at same
height as $\Xt_i$ or with higher height (effectively lower in lattice).

\begin{EXAMPLE}{}
For  constant  propagation, let  $c$  denote  any  constant. Then  the
longest strictly descending chain in \Lt\ is \mbox{$\Tt\sqsupset
c\sqsupset \Bt$} with heights of elements $(0,1,2)$ respectively.
For some \mbox{$\en_i^v\rightarrow\beta_j^u$}, condition
(\ref{cond:monotonicity.entity}) can only be violated if $\langle v, u
\rangle$ are $\langle c, \Tt \rangle, \langle \Bt, c \rangle$
or $\langle \Bt, \Tt \rangle$. However this is not possible
implying that constant propagation has monotonic entity
dependence. For faint variables analysis, the corresponding chain is
\mbox{$\Tt\sqsupset\Bt$}. It can be easily verified that the entity 
dependence is monotonic.
\end{EXAMPLE}

\ 

\begin{THEOREM}{thm:improvement}
For a data  flow framework with monotonic entity  dependence, the term
${\displaystyle\sum_{n=1}^{m}}  {\dd{\gamma_n\r\gamma_n}}$ in Equation
(\ref{eq:Delta.non-overlapping})               reduces              to
$\Ht\times\max(\weight(\gamma_n\r\gamma_n))$.
\end{THEOREM}

\begin{PROOF}
In path structure (\ref{eq:path.structure}), the multiplication factor
\Ht\ {in \dd{\gamma_n\r\gamma_n} (Equation~\ref{eq:path.cyclic})} is
required only if the flow is
\mbox{$\gamma_n^{\,v}\r\gamma_n^{\;u}$} such that \mbox{$ht(u) > ht(v)$}.
Due to monotonic entity dependence, the values computed in each cycle
will get progressively limited and instead of \Ht\ changes in each
cycle, the changes will get distributed over $m$ cycles. Let the
number of changes in cycle $\gamma_n\r\gamma_n$ be $c_n$. Then we need
to maximize the term
\begin{eqnarray*}
      {\displaystyle\sum_{n=1}^{m}}
      c_n\times\weight(\gamma_n\r\gamma_n)
\mbox{ such that }
      {\displaystyle\sum_{n=1}^{m}}
      c_n = \Ht
      \label{term:maxima}
\end{eqnarray*}
The constraint on $c_n$ defines a hyperplane and a maximum occurs 
at some extreme point of the plane. 
Hence the maximum value of the summation is
\(
 \Ht\times\max(\weight(\gamma_n\r\gamma_n))
\).
\hfill$\Box$
\end{PROOF}

\newcommand{\myspace}{\raisebox{-.15cm}{\rule{0cm}{.5cm}}}

\begin{figure*}[t]
\begin{center}
\begin{tabular}{p{9cm}}
\scalebox{.95}{
\begin{tabular}{|l|@{\ }c@{ }|@{\ }c@{ }|@{\ }c@{ }|@{\ }c@{ }|@{\ }c@{ }|@{\ }c@{ }|@{\ }c@{ }|}
\hline 
\raisebox{-.15cm}{\rule{0cm}{.5cm}} Benchmark & $\#F$ & $\#V$ &
$\;\depth\;$ & $\;\delta\;$ & $B_1$ &
$B_2$ & $I$\\ \hline \hline
\myspace arith\_coder & \ 29\  & \ 40.90\  & \ 1.31\  & \ 1.83\  & \ 117.76\  &
\ 4.14\  & \ 2.90\  \\ \hline
\myspace 186.crafty & 38 & 56.39 & 1.39 & 2.08 & 177.84 & 4.47 & 2.87 \\ \hline
\myspace whetstone & 2 & 9.00 & 1.00 & 1.50 & 19.00 & 3.50 & 2.50 \\ \hline
\myspace kexis & 11 & 33.27 & 1.09 & 1.55 & 76.45 & 3.64 & 2.55 \\ \hline
\myspace 175.vpr & 157 & 37.81 & 1.53 & 2.09 & 137.83 & 4.62 & 3.19 \\ \hline
\myspace 181.mcf & 20 & 25.65 & 1.35 & 1.50 & 69.30 & 3.85 & 2.30 \\ \hline
\myspace 197.parser & 213 & 11.78 & 1.37 & 1.27 & 39.97 & 3.64 & 2.53 \\ \hline
\myspace 164.gzip & 56 & 36.54 & 1.36 & 2.14 & 111.25 & 4.50 & 2.64 \\ \hline
\myspace gsm & 32 & 29.97 & 1.16 & 1.59 & 73.25 & 3.75 & 2.59 \\ \hline
\myspace 300.twolf & 110 & 46.26 & 1.67 & 1.83 & 160.58 & 4.50 & 2.94 \\ \hline
\myspace 256.bzip2 & 25 & 36.00 & 1.52 & 2.16 & 135.96 & 4.68 & 2.92 \\ \hline
\myspace 254.gap & 401 & 43.10 & 1.27 & 2.20 & 124.63 & 4.48 & 2.63 \\ \hline
\end{tabular}} \\  \\
(a) Summary of complexity computation for various benchmarks:
  $\#F$ is the number of functions, $\#V$ is the average number of
  variables. All other values are average values: $B_1$ is the
  predicted number of iterations using $1+\depth \times H$, $B_2$ is
  the predicted number of iterations using $1+\delta+\depth$, $I$ is
  the actual  number of iterations. 
\end{tabular}\hfill
\begin{tabular}{@{}p{6.5cm}}
\includegraphics[height=4.5cm,width=6.5cm]{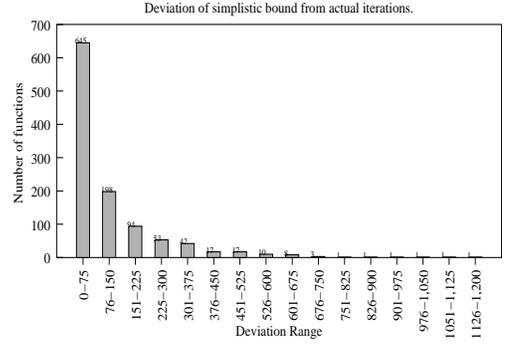}\\
{\hspace*{-.2cm}\small (b) Deviation of $B_1$ from actual number of iterations 
\raisebox{-.4cm}{\rule{0cm}{.2cm}}}\\ 
\includegraphics[height=4.5cm,width=6.5cm]{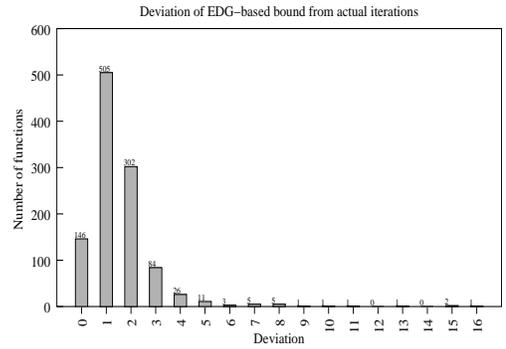}
{\small (c) Deviation of $B_2$  from actual number of iterations}
\end{tabular}
\end{center}
\caption{Empirical measurements}
\label{fig:table}
\rule{\textwidth}{.7pt}
\end{figure*}

\begin{THEOREM}{thm:main}
A round robin data flow analysis of an instance \idff\ of a data flow
framework \dff\ would converge in at most \mbox{$1+\delta+\depth$}
iterations.
\end{THEOREM}

\begin{PROOF}
The  first change  in  every $\en_0$  must  happen in  the very  first
iteration  because it  does  not  depend on  any  other entity.   From
Equation  (\ref{eq:degree.def}),  the  maximum  number  of  iterations
required for  computing the final value  of the data  flow property of
any entity  $\en_i$ is \mbox{$1+\delta$}.  However, it  is computed at
statement $i$ and  also needs to be propagated  to those statements in
the  \CFG\ for  which this  entity may  not occur  in \fuse.  This may
requires at most $\depth$ additional iterations taking the bound to
\mbox{$1+\delta+\depth$}.\hfill$\Box$
\end{PROOF}

\begin{COROLLARY}
For bit vector frameworks, the bound in Theorem~(\ref{thm:main})
reduces to \mbox{$1 + \depth$}.  
\end{COROLLARY}

\begin{PROOF}
For bit vector frameworks, \EDG\ does not contain any edges. Hence
\mbox{$\delta = 0$}. \hfill $\Box$
\end{PROOF}

\begin{EXAMPLE}{}
In      the      \EDG\      for      constant      propagation      in
Figure~\ref{fig:example.edgs}(a), \mbox{$\en_0=d_1$}. Since $\Ht = 2$,
$\dd{\en_0\r\beta_j}$  for each  $\beta_j$ in  $\{w_1, z_7,  y_6, x_5,
w_8\}$ is $\{0, 6, 6, 6,  6\}$ respectively. Hence $\delta = 6$. Depth
of the  \CFG\ is $\depth=3$.  Thus complexity bound  $(1+\delta+d)$ is
$10$.  This compares  well with the actual number  of iterations which
is 9.

In     the     \EDG\    for     faint     variables    analysis     in
Figure~\ref{fig:example.edgs}(b), \mbox{$\en_0=a_2$}. Since $\Ht = 1$,
$\dd{\en_0\r\beta_j}$  for each  $\beta_j$ in  $\{x_2, y_5,  z_6, w_7,
x_8\}$ is  $\{0,6,6,6,6\}$ respectively.   Hence $\delta =  6$.  Using
$\depth =  3$, we  get the overall  complexity bound $10$.  The actual
number of iterations is 7.
\end{EXAMPLE}

\section{Empirical Measurements}
\label{sec:data}

We have  implemented the complexity  computation and round  robin data
flow  analysis for  constant propagation  as well  as  faint variables
analysis     to    measure    the     precision    of     the    bound
\mbox{$1+\delta+\depth$}   for  practical  programs.    Our  prototype
implementation   uses    XSB   Prolog\footnote{Available   from   {\tt
http://xsb.sourceforge.net}}.   The input  is obtained  by translating
gimple IR produced by gcc version 4.0.0.

We have  tested our implementation on SPEC-2000  C benchmark programs,
whetstone and digital signal  processing benchmarks such as gsm. Since
the  implementation is  restricted to  intraprocedural level,  we make
conservative  assumptions  for  values  of global  variables  used  in
expressions.  Since  our  goal  is  not to  use  the  information  for
transformation but to measure the  precision of our bound, aliases are
ignored. We  have also computed the  simplistic bound (Section~\ref{})
$1+\depth\times H$ where $H = 2\times|V|$ for constant propagation.

Figure~\ref{fig:table}(a)   summarizes   the   results  for   constant
propagation presenting averages of various numbers. The table excludes
the information  of acyclic programs  since for acyclic  programs, the
bounds  are trivially  1.  Clearly,  our bound  is very  close  to the
actual number of  iterations.  Figures~\ref{fig:table}(b) and (c) plot
the number of functions against  the deviations of the bounds from the
actual     number     of     iterations.    Clearly,     the     bound
\mbox{$1+\delta+\depth$} has very small deviation (0 to 2) for a large
number ($87.11\%$) of programs. The bound $1+\depth\times H$ has large
deviations for most of the programs. In particular, small deviation (0
to 2) is  found in only $1.28\%$ cases. The  deviations of both bounds
is 0 for acyclic program and we have excluded them.

\section{Conclusions and Future Work}

Non-separability  of   data  flow  framework  is   a  dominant  factor
influencing  the complexity of  round robin  data flow  analysis.  The
existing theory accounts for lattice height and loop closure bounds in
determining complexity bounds. However,  it fails to capture the exact
role played  by non-separability.  This paper proposes  the concept of
degree of dependence which is a more precise measure since (a) It uses
the height of  the component lattice instead of  overall lattice while
considering  the cumulative effect  only for  interdependent entities,
and  (b) It  distinguishes  the  cyclic \CFG\  paths  from the  cyclic
dependences.   Apart  from  precision, these  distinctions  facilitate
generality by  placing separable  and non-separable frameworks  on the
same continuous  spectrum. This provides a uniform  explanation of the
phenomena  observed in  a  large  class of  practical  instances of  a
variety of data flow frameworks.

We    would    like    to     extend    this    work    to    point-to
analysis~\cite{emami-ip-points-to,aditya-hetero}  of C  and  C++ where
new entity  dependences are discovered during analysis  due to pointer
indirections.  Yet another interesting  direction of future work is to
explore the  use of entity  dependence graph for performing  data flow
analysis.


\subsection*{Acknowledgments}
We wish to thank Amitabha Sanyal, Supratim Biswas and Amey Karkare for
their useful comments.  Garima Lahoti helped in the prototype
implementation.

\bibliography{mybiblio}

\end{document}